\documentclass[runningheads]{waica}

\usepackage[T1]{fontenc}
\usepackage[utf8]{inputenc}
\usepackage{graphicx}
\usepackage{amsmath,amssymb}
\usepackage{booktabs}
\usepackage{multirow}
\usepackage{array}
\usepackage{xcolor}
\usepackage{url}
\usepackage{cite}
\usepackage{float}
\usepackage{enumitem}

\newcolumntype{L}[1]{>{\raggedright\arraybackslash}p{#1}}
\newcolumntype{C}[1]{>{\centering\arraybackslash}p{#1}}

\begin{document}

\title{RiverONE: Generating Knowledge-Intensive \\ VLM by Simulated Quantum Machines}

\author{Xindian Ma, Xinyu Long, Yefei Zhang, Yanchen Liu, Xianghao Li, Yufu Wen, \\Yike Hu, Yuedong Zhu, Zeyang Ma, Wen Qin,  Yikun Wang,Peng Yang\\ Monan Wang, Teng Yu}
\authorrunning{X. Ma et al.}
\institute{Thewake RiverYtz Lab\\ research@thewakesystems.com}

\maketitle

\begin{abstract}
Quantum computing provides a powerful paradigm for representing and transforming high-dimensional information through superposition, entanglement, and measurement-induced nonlinear features. While current quantum hardware is not yet practical for direct large-scale vision-language model (VLM) inference, simulated quantum computation can be used during model construction to generate structured parameters for compact classical AI systems. We build RiverONE, a lightweight vision-language model for quantum calibration plot understanding, using simulated quantum computation. It employs a specialized visual encoder and an InternVL-based language backbone. To compensate for compression-induced information loss, we introduce quantum-generated parameters, which are materialized as classical tensors after training. This allows RiverONE to run entirely on classical GPUs at inference time, with no quantum hardware or runtime quantum simulation. With approximately 1.9 billion parameters, RiverONE achieves at least 95\% of the performance of NVIDIA Ising Calibration 1 on quantum calibration plot understanding tasks while using less than 10\% of its parameter count. These results suggest that simulated quantum computation can serve as a practical construction-stage mechanism for building lightweight, knowledge-intensive scientific VLMs. Our code is available at https://github.com/THeWakeSystems/RiverOne.

\end{abstract}

\section{Introduction}

Quantum computing has long been studied for its ability to represent and manipulate information in ways that differ fundamentally from classical computation. For modern AI, this suggests a practical possibility: quantum circuits may provide structured parameterizations for building smaller neural models, even when the final deployed system remains entirely classical. Through data encoding, parameterized unitary evolution, entanglement, and measurement, quantum circuits define nonlinear transformations over high-dimensional representations~\cite{biamonte2017quantum,cerezo2021vqa,havlicek2019quantum,schuld2021encoding}. The expressive behavior of these transformations depends on the encoding scheme, circuit depth, entanglement pattern, measurement observables, and repeated data re-uploading~\cite{schuld2021encoding,perezsalinas2020data,mitarai2018quantum}. Recent hybrid architectures further support this construction-stage view by using variational quantum circuits to generate classical neural-network weights during training while keeping deployment fully classical~\cite{qi2025vqcmlpnet}.

Meanwhile, large language models (LLMs) and vision-language models (VLMs) have become general-purpose engines for reasoning, coding, scientific question answering, and multimodal understanding~\cite{vaswani2017attention,brown2020language,radford2021clip,alayrac2022flamingo,li2023blip2,liu2023llava,wang2025internvl3_5}. Their capabilities are closely tied to scale: empirical scaling laws show that performance improves predictably with model size, data size, and training compute, while compute-optimal training further couples larger models with larger training corpora~\cite{kaplan2020scaling,hoffmann2022training}. However, scaling also brings billions or tens of billions of parameters, large memory footprints, and expensive inference infrastructure. These costs are especially problematic for scientific AI systems that must operate near instruments, laboratories, or edge devices. This has motivated efficient model construction methods such as knowledge distillation, pruning, post-training quantization, activation-aware quantization, low-rank adaptation, and additive codebook compression~\cite{hinton2015distilling,han2015deepcompression,frantar2023gptq,xiao2023smoothquant,lin2024awq,hu2022lora,aqlm}. We connect these two directions by using simulated quantum computation as a construction-stage parameter generator to compensate compressed scientific VLMs.

Our target setting is quantum device calibration, a knowledge-intensive scientific workflow in which researchers and engineers interpret plots such as Rabi oscillations, Ramsey fringes, resonance scans, spectroscopy heatmaps, decay traces, and readout-cluster distributions~\cite{krantz2019quantum,kjaergaard2020superconducting,qiskitexperiments,qcaleval2026}. These plots are not ordinary charts: their visual patterns encode physical information about frequency alignment, coherence, fit quality, parameter values, noise structure, and calibration actions. A useful calibration VLM must connect low-level visual evidence, such as oscillation contrast, envelope decay, peak displacement, fitting residuals, and cluster separation, with domain-specific conclusions about whether an experiment succeeded, whether a fitted curve is reliable, which parameter should be extracted, and what calibration step should follow.

QCalEval formalizes quantum calibration plot understanding as a VLM benchmark, covering calibration-relevant question types such as visual description, outcome classification, scientific reasoning, fit reliability, parameter extraction, and calibration diagnosis~\cite{qcaleval2026}. Unlike generic chart and plot benchmarks such as PlotQA, ChartQA, and DePlot~\cite{methani2020plotqareasoningscientificplots,masry2022chartqabenchmarkquestionanswering,deplot}, QCalEval requires models to connect subtle visual evidence with physical conclusions and calibration actions. This exposes a central challenge for scientific VLMs: large general-purpose models may provide strong visual-language reasoning, but practical calibration workflows require models that are both domain-aware and lightweight enough for local or low-latency deployment.

Model compression is a natural path toward deployable scientific VLMs, but it can remove the very capacity needed for calibration reasoning. On the visual side, MiniViT-style weight multiplexing compresses Vision Transformers by sharing weights across consecutive Transformer blocks while using lightweight transformations and attention distillation to preserve layer diversity~\cite{minivit}. On the language side, additive codebook quantization methods such as AQLM approximate full-precision LLM weights through sums of learned codebook entries, enabling aggressive low-bit compression with reduced memory footprint~\cite{aqlm}. These methods reduce model size, but quantum calibration plot understanding is highly sensitive to compression-induced information loss: weakened layer-specific attention may hurt fit assessment, while degraded language-side reasoning may affect parameter extraction and calibration diagnosis.

We introduce RiverONE, a lightweight VLM for quantum calibration plot understanding that uses simulated quantum computation as a construction-stage compensation mechanism. RiverONE combines an InternVL3.5-based language backbone~\cite{wang2025internvl3_5} with a calibration-specialized visual encoder derived from the SigLIP2-style visual representation learning~\cite{vit,siglip2}. The base model is compressed through MiniViT-style visual weight sharing and additive codebook quantization of the language backbone. To compensate for the capacity lost during visual compression, we propose QGP, a simulated variational quantum circuit-based generator that produces layer-specific compensation parameters for shared visual Transformer blocks.

The key idea of QGP is to connect quantum computation with LLM/VLM compression without requiring quantum inference. During model construction, compressed shared weights and lightweight layer-level conditioning signals are encoded into a simulated variational quantum circuit. The circuit produces measurement features that are mapped into compact compensation parameters, such as attention perturbation coefficients, residual gates, or low-rank corrections. After training, these generated parameters are materialized into ordinary classical tensors and saved in the model checkpoint. Thus, the final RiverONE model runs entirely on classical GPUs, requiring no quantum hardware, no runtime circuit execution, and no quantum simulation overhead. This follows the construction-stage philosophy of hybrid quantum-classical parameter generation, where quantum circuits are used to generate classical neural parameters during training while deployment remains classical~\cite{qi2025vqcmlpnet}.

We study whether simulated quantum parameter generation can serve as a useful inductive bias for compact scientific VLM construction. The variational quantum circuit readout defines a structured nonlinear map over encoded weight representations: after amplitude encoding and unitary mixing, measurement probabilities contain cross-entry interactions among encoded amplitudes~\cite{havlicek2019quantum,schuld2021encoding,perezsalinas2020data}. This motivates QGP as a compact nonlinear generator for compression compensation. We empirically test whether this quantum-structured generator recovers calibration-specific reasoning ability more effectively than parameter-matched classical compensation baselines.

We evaluate RiverONE on QCalEval against NVIDIA Ising Calibration 1, general-purpose VLMs, and InternVL3.5 models at multiple scales~\cite{qcaleval2026,wang2025internvl3_5}. Our experiments separate the effects of domain fine-tuning, visual compression, language compression, and compensation. In particular, we compare compressed RiverONE variants with classical MLP adapter, and the proposed QGP module under matched parameter budgets, focusing on calibration-sensitive categories such as fit reliability, parameter extraction, and diagnosis.

Our contributions are summarized as follows:
\begin{itemize}
    \item \textbf{A quantum-for-LLM construction framework for scientific VLM compression.}
    We formulate simulated quantum parameter generation as a construction-stage mechanism for building compact classical VLMs, using quantum-structured computation to compensate a compressed multimodal model without quantum inference.

    \item \textbf{RiverONE, a compact VLM for quantum calibration plot understanding.}
    We develop RiverONE for QCalEval-style calibration reasoning by combining a calibration-specialized visual encoder with an InternVL3.5-based language backbone under a deployment-oriented compression budget.

    \item \textbf{QGP, a simulated variational quantum circuit-based generator for compression compensation.}
    We introduce QGP to generate layer-specific compensation parameters for shared visual Transformer blocks. The generated parameters are materialized into classical tensors after training, so the final model remains a standard classical VLM at inference time.

\end{itemize}

\section{RiverONE}

Compared with NVIDIA Ising Calibration 1~\cite{qcaleval2026}, the RiverONE series achieves comparable calibration-plot understanding performance with substantially fewer parameters, demonstrating the effectiveness of our compression and compensation pipeline. In this section, we describe the construction of RiverONE in a step-by-step manner. Section~\ref{sec:base-architecture} first introduces the base RiverONE architecture, including the calibration-specialized visual encoder, the vision-language projector, and the language backbone. We then present the two main compression components: Section~\ref{com:vit_shared} describes the shared-Transformer mechanism used to compress the visual encoder, while Section~\ref{com:LLM_com} introduces the compression of the language model layers. Section~\ref{com:VQC_generation} further explains how VQC-based parameter generation compensates for compression-induced representational loss and enhances the compact model. Finally, Section~\ref{sec:train} details the post-training methods and construction procedures used to obtain the final RiverONE models.

\subsection{Base Architecture}
\label{sec:base-architecture}

\begin{figure}[h]
\centering
\includegraphics[width=0.95\linewidth]{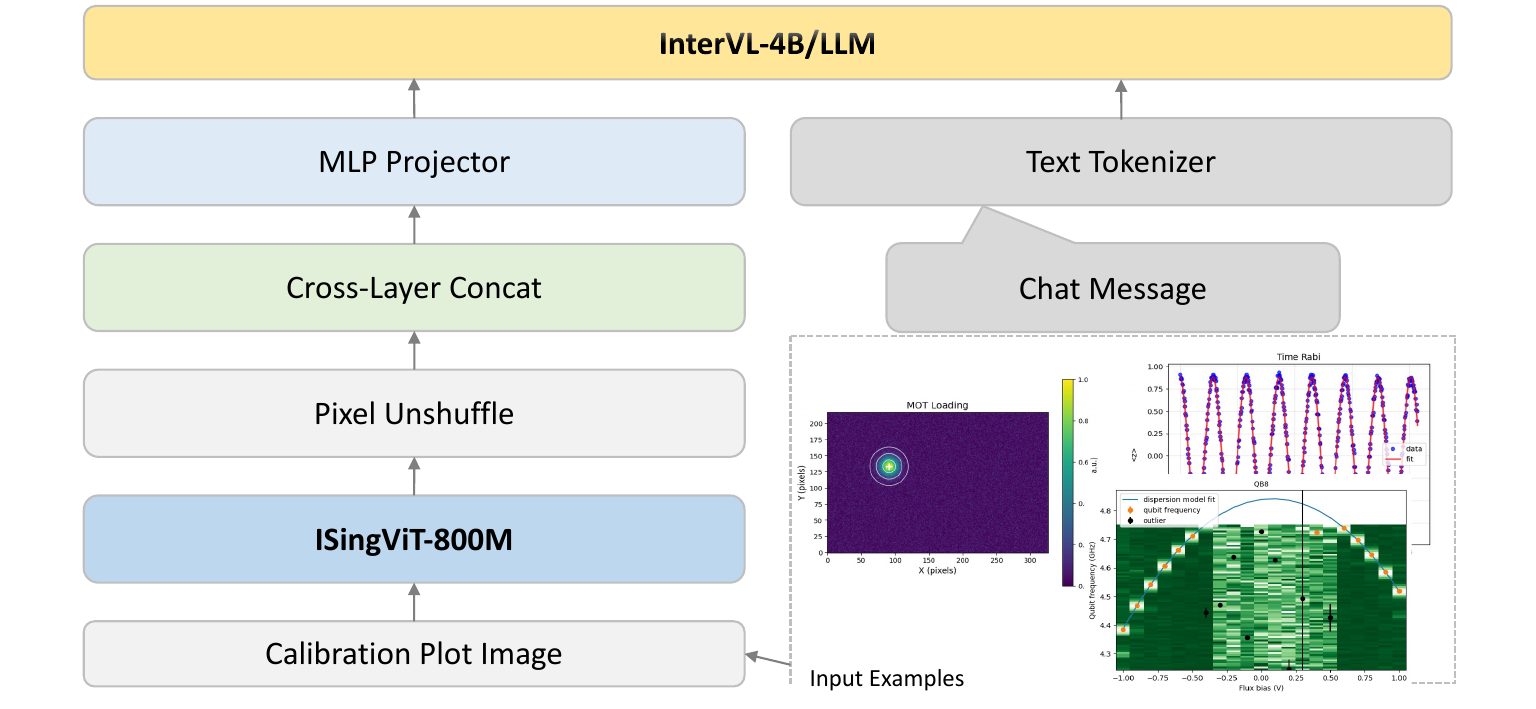}
\caption{Base RiverONE architecture. A calibration plot image is processed by \textsc{ISingViT}-800M, followed by pixel unshuffle, cross-layer concatenation, and an MLP projector. The projected visual embeddings are combined with tokenized chat messages and passed into the Large Language Model (LLM) of InternVL-4B.}
\label{fig:base_architecture}
\end{figure}

RiverONE is a vision-language model that processes a calibration plot image $x \in \mathbb{R}^{H \times W \times 3}$ and a text query $q$ to produce an autoregressive response $y = (y_1, y_2, \ldots, y_m)$. The model consists of four components: a visual encoder $E_{\phi}^{\mathrm{Ising}}$ parameterized by $\phi$, a pixel-unshuffle operator $U$, a cross-layer concatenation operator $C$, an MLP projector $P_{\psi}$ parameterized by $\psi$, and a language model $L_{\theta}$ parameterized by $\theta$. The overall architecture is illustrated in Figure~\ref{fig:base_architecture}.

\paragraph{\textbf{Visual encoding.}}
The input image $x$ is processed by IsingViT-800M, a domain-specialized Vision Transformer pre-trained on quantum calibration imagery. The encoder produces a sequence of layer-wise hidden states; we denote the full set of intermediate representations collectively as
\begin{equation}
    H = E_{\phi}^{\mathrm{Ising}}(x),
    \quad H \in \mathbb{R}^{L \times N \times d},
\end{equation}
where $L$ is the number of Transformer blocks, $N$ is the number of patch tokens, and $d$ is the hidden dimension.

\paragraph{\textbf{Pixel unshuffle and cross-layer concatenation.}}
Before projection, the visual representations undergo two operations that increase spatial and semantic richness. Pixel unshuffle $U$ rearranges spatial sub-pixels of the patch feature maps to increase the channel dimension while reducing spatial resolution, compressing spatially redundant information without discarding it:
\begin{equation}
    H^{u} = U(H), \quad H^{u} \in \mathbb{R}^{L \times N' \times (d \cdot r^{2})},
\end{equation}
where $r$ is the unshuffle factor and $N' = N / r^{2}$.
Cross-layer concatenation $C$ then merges representations from multiple Transformer layers along the channel dimension, allowing downstream modules to access features at different levels of abstraction simultaneously:
\begin{equation}
    H^{c} = C(H^{u}), \quad H^{c} \in \mathbb{R}^{N' \times (L \cdot d \cdot r^{2})}.
\end{equation}

\paragraph{\textbf{MLP projection.}}
The concatenated visual representation $H^{c}$ is mapped into the language model's embedding space by a two-layer MLP projector $P_{\psi}$:
\begin{equation}
    Z = P_{\psi}(H^{c}), \quad Z \in \mathbb{R}^{N' \times d_{\mathrm{lm}}},
\end{equation}
where $d_{\mathrm{lm}}$ is the hidden dimension of the language model. The projector serves as the sole trainable bridge between the visual and language streams and is learned end-to-end.

\paragraph{\textbf{Language Modeling.}}
The text query $q$ is tokenized and embedded separately. The projected visual tokens $Z$ and the text embeddings are concatenated and fed into the language model $L_{\theta}$, which is the 4B-parameter backbone of InternVL3\_5-4B. The model produces a conditional distribution over the next response token at each decoding step:
\begin{equation}
    p_{\theta}(y_t \mid y_{<t},\, x,\, q)
    = L_{\theta}(Z,\, q,\, y_{<t}).
\end{equation}
The response is generated autoregressively by sampling from this distribution.

\begin{figure}[h]
\centering
\includegraphics[width=0.95\linewidth]{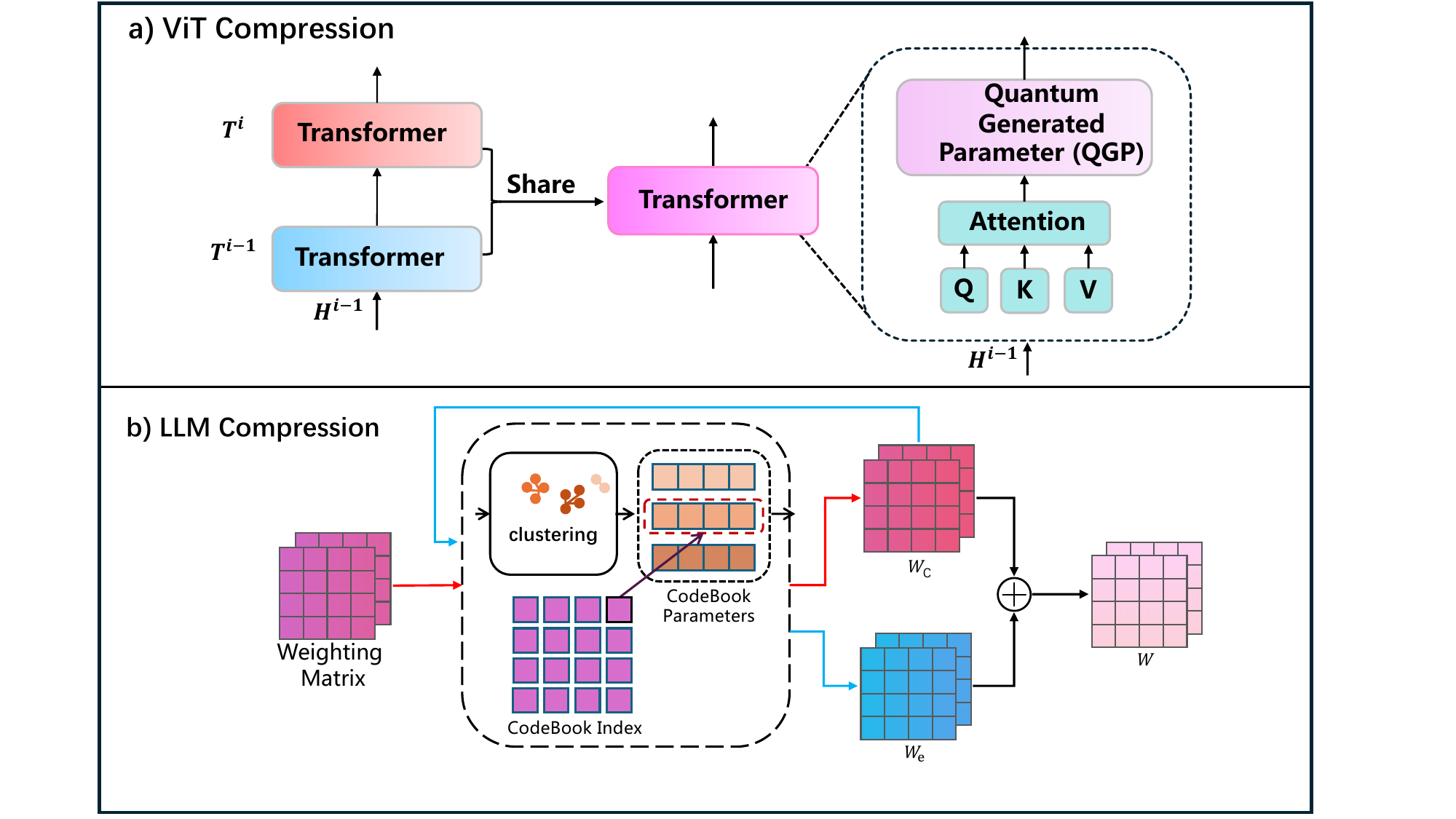}
\caption{Compression design of RiverONE. The visual encoder is compressed through Transformer block sharing, where shared attention blocks are compensated by Quantum-Generated Parameter (\textsc{QGP}). The LLM is compressed through the codebook quantization by Quantum Inspired.}
\label{fig:compression}
\end{figure}

\subsection{Visual compression with shared Transformers}
\label{com:vit_shared}
Vision Transformers~\cite{vit} stack many structurally similar blocks, which motivates cross-layer sharing and weight multiplexing strategies such as MiniViT~\cite{minivit}. In this work, we employ a variational quantum circuit to generate the parameters for constructing this linear transformation. Then, self‑attention distillation is leveraged to recover the representational capacity. Figure~\ref{fig:compression} (a) shows a schematic diagram of the parameter compression of this architecture.

To reduce the visual encoder cost, RiverONE compresses repeated Transformer computation through block sharing following this principle.
Let the original visual encoder contain $L$ Transformer blocks:
\begin{equation}
    T^1,\, T^2,\, \ldots,\, T^L.
\end{equation}
A compressed encoder replaces them with $K$ shared blocks and their associated per-layer weight transformations:
\begin{equation}
    \hat{T}^1,\, \hat{T}^2,\, \ldots,\, \hat{T}^K,
    \qquad K < L.
\end{equation}
A layer mapping $g : \{1,\ldots,L\} \rightarrow \{1,\ldots,K\}$ assigns each original layer index to one shared block. The hidden state at layer $i$ is then computed as
\begin{equation}
    H^i = \hat{T}^{g(i)}\!\left(H^{i-1};\, \phi^i\right),
\end{equation}
where $\phi^i$ denotes the lightweight per-layer transformation applied to the shared weights of block $g(i)$ before the forward pass, following the weight-multiplexing design of~\cite{minivit}.

Weight multiplexing reduces the total parameter count substantially, but it also constrains layer-specific representational diversity. In calibration plots, different layers capture qualitatively distinct forms of evidence: early layers respond to local curve smoothness and edge structure, intermediate layers localize peaks, fitted envelopes, and spectral features, and later layers encode higher-order patterns such as cluster separation and noise texture. Collapsing $L$ independent weight sets into $K$ shared ones limits the model's ability to maintain this hierarchy of calibration-relevant visual features. The per-layer transformation $\phi^i$ partially compensates for this diversity loss, but it operates on the shared weights globally and cannot recover fine-grained, layer-specific calibration structure. 
This motivates the targeted compensation mechanism, i,e., Quantum Generated Parameter(QGP) described in Section~\ref{com:VQC_generation}.

\subsection{Quantum Generation Parameter}

\begin{figure}[h]
\centering
\includegraphics[width=0.8\linewidth]{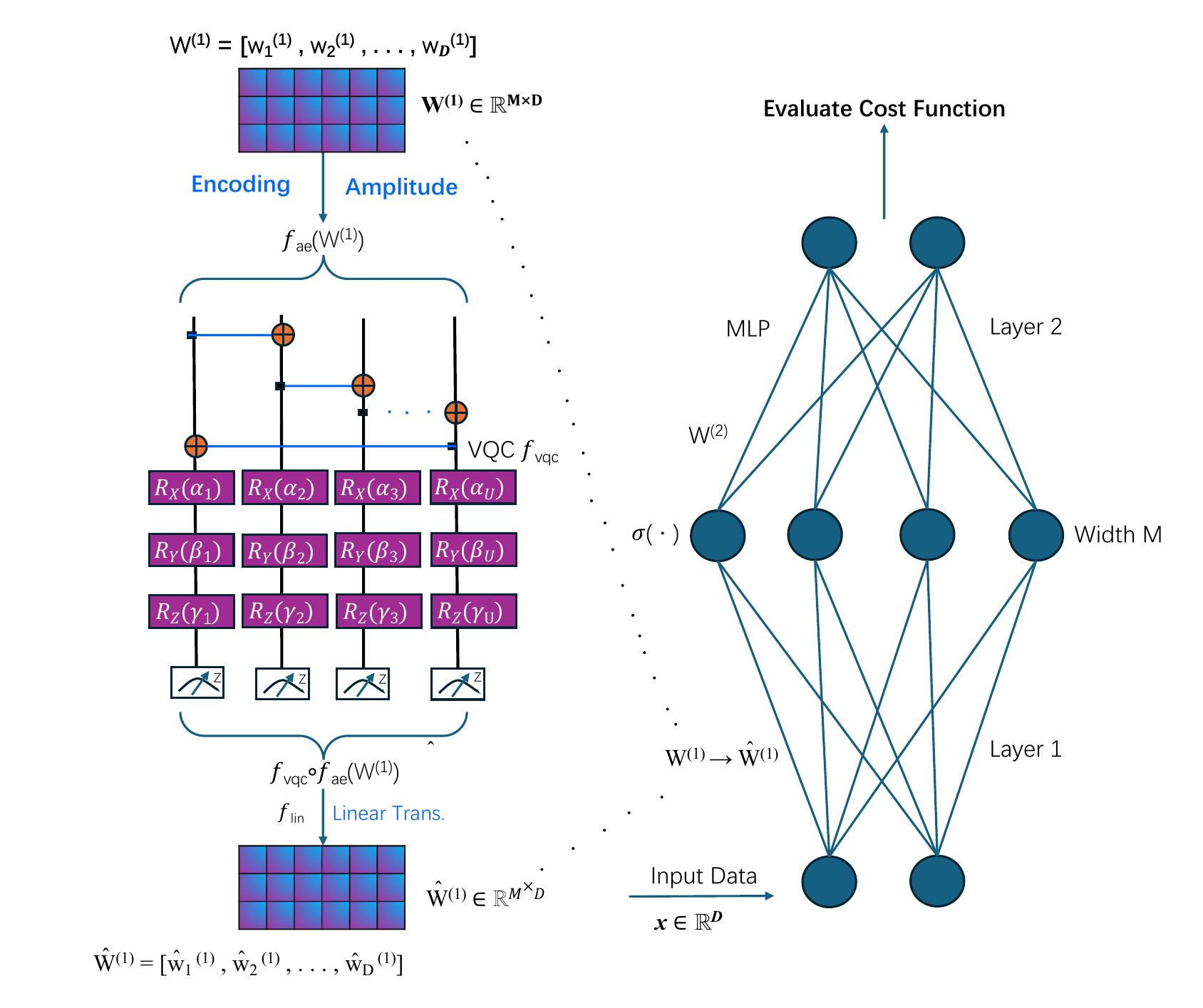}
\caption{\textbf{Overview of the \textsc{QGP} weight-transformation pipeline.} Layer‑specific weights are generated from a shared compressed matrix via amplitude encoding, VQC, and linear mapping. They are inserted into the encoder for end‑to‑end training and then materialized as classical tensors – quantum processing is only used during training, not inference.
}
\label{fig:qgp}
\end{figure}

\label{com:VQC_generation}
Weight multiplexing compresses the visual encoder by sharing attention weights across layers, but the resulting shared weights $\hat{W}_Q, \hat{W}_K, \hat{W}_V$ lose the layer-specific expressiveness required for fine-grained calibration reasoning. The \textsc{QGP} module addresses this by transforming the shared weights into layer-adapted versions using a simulated variational quantum circuit (VQC). Rather than adding a separate correction term, \textsc{QGP} directly produces a quantum-enhanced reparameterization of the shared weights for each selected attention block, as illustrated in Fig.~\ref{fig:qgp}.

A standard self-attention block at layer $i$ computes queries, keys, and values from the hidden state $H^{i-1} \in \mathbb{R}^{N \times d}$:
\begin{equation}
    Q = H^{i-1}W_Q, \qquad
    K = H^{i-1}W_K, \qquad
    V = H^{i-1}W_V,
\end{equation}
and produces the attention output
\begin{equation}
    A(H^{i-1}) = \mathrm{softmax}\!\left(\frac{QK^\top}{\sqrt{d}}\right)V.
\end{equation}
After weight multiplexing, $W_Q, W_K, W_V$ are replaced by shared compressed matrices $\hat{W}_Q, \hat{W}_K, \hat{W}_V$. \textsc{QGP} produces layer-specific reparameterized weights $\hat{W}_Q^i, \hat{W}_K^i, \hat{W}_V^i$ for each selected block $i$, so that the compensated attention becomes
\begin{equation}
    \hat{A}_i(H^{i-1})
    = \mathrm{softmax}\!\left(\frac{H^{i-1}\hat{W}_Q^i\,
      (H^{i-1}\hat{W}_K^i)^\top}{\sqrt{d}}\right)
      H^{i-1}\hat{W}_V^i.
\end{equation}

\paragraph{Amplitude encoding.}
The shared weight matrix $\hat{W}_Q \in \mathbb{R}^{M \times D}$ is encoded into a quantum state via amplitude encoding $f_{ae}$. The $MD$ scalar values are normalized and loaded as probability amplitudes over $n_q = \lceil \log_2(MD) \rceil$ qubits:
\begin{equation}
    |\psi_0\rangle = f_{ae}(\hat{W}_Q)
    = \sum_{j=0}^{2^{n_q}-1} c_j |j\rangle,
    \qquad c_j \in \mathbb{R},\quad \sum_j c_j^2 = 1.
\end{equation}
The same procedure is applied independently to $\hat{W}_K$ and $\hat{W}_V$.

\paragraph{Variational quantum circuit.}
The encoded state is processed by a parameterized ansatz $U(\alpha, \beta, \gamma)$ composed of alternating entanglement and rotation layers. Each rotation layer applies single-qubit $R_X$, $R_Y$, $R_Z$ gates to every qubit $u \in \{1, \ldots, n_q\}$:
\begin{equation}
    R(\alpha_u, \beta_u, \gamma_u)
    = R_X(\alpha_u)\, R_Y(\beta_u)\, R_Z(\gamma_u),
\end{equation}
where $R_\sigma(\theta) = e^{-i\theta \sigma / 2}$ for Pauli operator $\sigma \in \{X, Y, Z\}$. Between rotation layers, a chain of CNOT gates entangles adjacent qubits, coupling their probability amplitudes across the full $n_q$-qubit register. The full circuit prepares the state
\begin{equation}
    |\psi(\alpha, \beta, \gamma)\rangle
    = U(\alpha, \beta, \gamma)\,|\psi_0\rangle,
\end{equation}
and each basis state $|j\rangle$ is assigned a measurement probability
\begin{equation}
    p_j = \bigl|\langle j \mid \psi(\alpha, \beta, \gamma)\rangle\bigr|^2.
\end{equation}

\paragraph{Linear readout and weight reconstruction.}
The measurement probability vector $(p_0, \ldots, p_{2^{n_q}-1})$ is mapped back to a matrix of the same shape as the original shared weight via a learned linear transformation $f_{lin}$:
\begin{equation}
    \hat{W}_Q^i = f_{lin}\bigl(f_{vqc}(f_{ae}(\hat{W}_Q;\, \alpha, \beta, \gamma))\bigr)
    \in \mathbb{R}^{M \times D}.
\end{equation}
The full pipeline for block $i$ is therefore
\begin{equation}
    \hat{W}^i = f_{lin} \circ f_{vqc} \circ f_{ae}(\hat{W};\, \alpha^i, \beta^i, \gamma^i),
\end{equation}
where $\alpha^i, \beta^i, \gamma^i$ are layer-specific circuit parameters and $\hat{W}$ denotes each of $\hat{W}_Q, \hat{W}_K, \hat{W}_V$ in turn. The entanglement structure of the VQC allows the measurement distribution to encode correlations across the full weight matrix that a same-size classical adapter cannot represent, providing the additional expressive capacity needed to recover calibration-specific attention structure.

\paragraph{Classical deployment.}
\textsc{QGP} operates exclusively during the compression and fine-tuning stage, where the VQC is executed on a classical simulator. Once training converges, the reparameterized weight matrices $\hat{W}_Q^i, \hat{W}_K^i, \hat{W}_V^i$ are materialized into static classical tensors and saved as part of the model checkpoint. Inference therefore requires no quantum circuit execution: RiverONE runs as a standard classical model on conventional GPUs, with the quantum-derived weights indistinguishable from any other trained parameter tensor.

\subsection{Language Model Compression by Quantum Inspired}
\label{com:LLM_com}

A quantum pure state $|\psi\rangle$ is a superposition of basis states: $|\psi\rangle = \sum_j c_j |j\rangle$, where each $|j\rangle$ contributes independently and the full state is their weighted sum. Additive codebook quantization~\cite{aqlm} follows the same structural principle applied to weight matrices. A full-precision weight matrix $W \in \mathbb{R}^{M \times D}$ is approximated as a superposition of $J$ learned codebook contributions, each acting as a basis element in the space of possible weight configurations:
\begin{equation}
    W \approx \widehat{W} = \sum_{j=1}^{J} C_j[I_j].
    \label{eq:aqlm}
\end{equation}
Here $C_j \in \mathbb{R}^{|\mathcal{C}| \times D}$ is the codebook size, and $I_j \in \{1, \ldots, |\mathcal{C}|\}^M$ stores the discrete row-wise assignment into that codebook, so $C_j[I_j] \in \mathbb{R}^{M \times D}$ selects one row of $C_j$ for each output dimension. Each codebook $C_j$ forms a distinct basis that captures one specific mode of the weight distribution, analogous to the way orthogonal basis states in a quantum superposition each contribute a separable, non‑redundant component to the overall state. Fig.~\ref{fig:compression}. (b) shows an example with $J=2$.

\paragraph{Additive structure as quantum-inspired superposition.}
Standard scalar quantization represents each weight independently at low bit-width, discarding inter-weight structure. By contrast, the additive decomposition in Equation~(\ref{eq:aqlm}) encodes $W$ as a composition of $J$ codebook bases, where each basis captures global correlations across the output dimension. Increasing $J$ improves the approximation by incorporating additional basis contributions, much like adding higher‑order terms in a quantum state expansion reduces truncation error. This makes the approximation quality tunable: a single codebook ($J=1$) gives a coarse but memory-minimal representation, while $J > 1$ recovers finer weight structure at the cost of storing additional index tensors. For the language backbone of RiverONE, which must retain domain-specific calibration reasoning within a 1.5B-parameter budget, the multi-basis construction provides a more faithful reconstruction of the original weight distribution than single-codebook or uniform scalar schemes of the same storage cost~\cite{aqlm}.

\paragraph{Application to the language backbone.}
RiverONE applies additive codebook quantization to all linear projections of the Qwen3-based language backbone. The codebooks $\{C_j\}_{j=1}^{J}$ and index tensors $\{I_j\}_{j=1}^{J}$ are jointly optimized by minimizing the layer-wise weight reconstruction error on a calibration dataset, following the post-training quantization protocol of AQLM~\cite{aqlm}. At inference, $\widehat{W}$ is reconstructed on-the-fly from the stored codebooks and indices; no full-precision weight tensor needs to reside in GPU memory. This reduces the memory footprint of the language backbone by over one order of magnitude relative to BFloat16 storage while preserving the multi-token, domain-specific reasoning required by QCalEval tasks.

\subsection{Training objective}
\label{sec:train}
Post-training of RiverONE proceeds through two sequential optimization objectives. We first apply Supervised Fine-Tuning (SFT) to establish a strong base for calibration-domain instruction following, then apply Mixed Preference Optimization (MPO) to further align model outputs with expert-preferred responses.

\paragraph{\textbf{Supervised Fine-Tuning (SFT).}}
SFT is conducted in two consecutive stages that differ in the supervision signal and the expected generalization regime.

\textit{Phase 1: In-context learning SFT.}
The first phase trains RiverONE on demonstration style examples in which the input includes one or more reference question–answer pairs alongside the target question. 
This in-context format teaches the model to exploit provided exemplars for calibration reasoning tasks, building the initial domain grounding needed for specialized visual understanding.

\textit{Phase 2: Zero-shot SFT.}
The second phase removes all in-context demonstrations and fine-tunes the model on stand-alone question–answer pairs. This forces the model to internalize calibration knowledge rather than rely on exemplars. Given a supervised dataset $\mathcal{D} = \{(x_i, q_i, y_i)\}_{i=1}^{N}$, both phases minimize the standard negative log-likelihood objective:
\begin{equation}
    \mathcal{L}_{\mathrm{SFT}}
    = -\frac{1}{N}\sum_{i=1}^{N}\sum_{t=1}^{T_i}
    \log p_{\theta}(y_{i,t} \mid y_{i,<t},\, x_i,\, q_i).
    \label{eq:sft}
\end{equation}

\paragraph{\textbf{Mixed Preference Optimization (MPO).}}
Following SFT, we apply Mixed Preference Optimization (MPO) within an offline reinforcement learning framework~\cite{internvl_mpo}. MPO is applied to preference pairs $(y^+, y^-)$, where $y^+$ is more visually grounded or scientifically correct than $y^-$. The MPO training objective combines three weighted loss terms:
\begin{equation}
    \mathcal{L}_{\mathrm{MPO}}
    = \mathcal{L}_{\mathrm{pref}}
    + \lambda_{\mathrm{qual}}\,\mathcal{L}_{\mathrm{qual}}
    + \lambda_{\mathrm{gen}}\,\mathcal{L}_{\mathrm{gen}},
    \label{eq:mpo}
\end{equation}
where each term addresses a distinct aspect of output quality:

\begin{itemize}
    \item \textbf{Preference loss} $\mathcal{L}_{\mathrm{pref}}$ (DPO loss): contrasts accepted and rejected responses relative to a frozen reference policy $\pi_{\mathrm{ref}}$,
    \begin{equation}
        \mathcal{L}_{\mathrm{pref}}
        = -\mathbb{E}\log\sigma\!\Bigg(\beta\Big[
            \log\frac{\pi_\theta(y^+\mid x,q)}{\pi_{\mathrm{ref}}(y^+\mid x,q)}
            - \log\frac{\pi_\theta(y^-\mid x,q)}{\pi_{\mathrm{ref}}(y^-\mid x,q)}
        \Big]\Bigg).
    \end{equation}

    \item \textbf{Quality loss} $\mathcal{L}_{\mathrm{qual}}$ (BCO loss): provides a binary correctness signal on individual responses, enabling efficient use of unpaired quality annotations.

    \item \textbf{Generation loss} $\mathcal{L}_{\mathrm{gen}}$ (standard LM loss): retains the language modeling objective on accepted responses to prevent reward hacking and maintain fluency.
\end{itemize}

\paragraph{\textbf{Full training objective.}}
The complete post-training objective is the sum of the SFT loss (Equation~\ref{eq:sft}), the MPO composite loss (Equation~\ref{eq:mpo}), and a regularization term $\mathcal{L}_{\mathrm{reg}}$ that applies norm regularization to the QGP compensation parameters:
\begin{equation}
    \mathcal{L}
    = \mathcal{L}_{\mathrm{SFT}}
    + \mathcal{L}_{\mathrm{MPO}}
    + \lambda_{\mathrm{reg}}\,\mathcal{L}_{\mathrm{reg}}.
\end{equation}
SFT is optimized first; MPO is applied subsequently on the SFT-converged checkpoint. This sequential schedule ensures that preference optimization refines an already calibration-capable model rather than competing with basic instruction following from the start.

\section{Experiments}
Here, we first describe in Section~\ref{sec:dataconstruct} the construction process of our multimodal scientific image reasoning dataset and the final composition of the training data. Section~\ref{sec:baselines} introduces the VLMs we compare against, including both open‑source and closed‑source models. In Section~\ref{sec:mainresult}, we present the comparison of main results and discuss the experimental findings. Section 3.4 provides ablation studies that analyze the effectiveness of our quantum‑inspired and quantum‑circuit‑based parameter compression method during model compression.

\subsection{Dataset Construction}
\label{sec:dataconstruct}
Our dataset construction follows a two-stage pipeline: Supervised Fine-Tuning (SFT) dataset construction and Mixed Preference Optimization (MPO) dataset construction. The SFT stage builds general chart, plot, and science-figure understanding, while the MPO stage provides targeted preference data for improving QCalEval-style scientific reasoning. Table~\ref{tab:data_summary} provides an overview of the dataset used in our two-stage construction pipeline.  

\begin{table}[h]
\centering
\caption{Summary of constructed datasets used in different training stages.}
\label{tab:data_summary}
\small
\setlength{\tabcolsep}{4pt}
\begin{tabular}{p{0.10\linewidth} p{0.30\linewidth} p{0.20\linewidth} p{0.33\linewidth}}
\toprule
Stage & Data source & Size & Purpose \\
\midrule
SFT & ChartQA & 5,000 samples & General chart understanding, including axis reading, value comparison, and trend identification. \\
SFT & PlotQA & Additional samples & Plot-based visual question answering and general plotted-data interpretation. \\
SFT  & ScienceQA & Filtered subset & Science-oriented visual evaluation. Samples are filtered by an external model API to keep image-dependent science questions. \\
SFT & Qiskit Calibration Drift & Constructed subset & Quantum hardware drift QA data based on time-series drift curves and environmental correlation heatmaps. \\
MPO &  Qwen3.5-VL Construct & 4,326 samples & Calibration-specific expansion for scientific reasoning, fit reliability, parameter extraction, and diagnosis. \\
MPO & ChartQA & 500 retained from 5,000 candidates & High-quality chart reasoning samples for preserving general visual understanding during MPO training. \\
MPO & MMPR-v1.2 & Filtered subset & Science-related multimodal preference data, including scientific plots, experimental curves, and fitting-related figures. \\
\bottomrule
\end{tabular}
\end{table}

\paragraph{Supervised Fine-Tuning Datasets}.
The SFT dataset is designed to build general chart, plot, and science-figure understanding.
\begin{enumerate}[leftmargin=*]
    \item \textbf{ChartQA}~\cite{masry2022chartqabenchmarkquestionanswering} \textbf{and PlotQA}~\cite{methani2020plotqareasoningscientificplots}. ChartQA and PlotQA provide general chart-based visual question answering samples, including bar charts, line charts, scatter plots, and other statistical figures. We process 5,000 ChartQA samples together with additional PlotQA samples, remove invalid or malformed examples, and convert the remaining data into a unified multimodal instruction format with an image, a question, and a reference answer.
    \item \textbf{ScienceQA filtering}~\cite{lu2022learn}. ScienceQA is used to construct a science-oriented evaluation subset. We first remove examples without valid images, and then use Qwen3.5-122B-A10B to select image-dependent science questions, including scientific diagrams, experimental setups, maps, charts, and other visual reasoning cases.
    \item \textbf{Qiskit Calibration Drift}~\cite{qiskit_calibration_drift}. Qiskit Calibration Drift is used to construct quantum hardware drift QA data. The raw records are grouped by device, qubit, and drift property. We generate 14-day time-series drift curves and 30-day environmental correlation heatmaps, and then create QA pairs about drift statistics, temporal trends, system stability, and environmental factors affecting calibration drift.
\end{enumerate}
\paragraph{Mixed Preference Optimization Datasets.} The MPO dataset is designed to provide targeted preference signals for QCalEval-style scientific reasoning. 

\begin{enumerate}[leftmargin=*]

    \item \textbf{Qwen3.5-VL Construct.} We utilize the image data from the QCalEval work for data augmentation, and the augmented data is used for model training; the original benchmark data is not used for model training. For each selected image, Qwen3.5-122B-A10B~\cite{qwen35} generates three new question-answer pairs based on the visual content and the original QCalEval task context. The generated questions cover calibration-related abilities such as visual description, scientific reasoning, fit reliability, parameter extraction, and calibration diagnosis. We then convert the generated samples into the MPO preference format, where each example contains \texttt{id}, \texttt{image}, \texttt{question}, \texttt{chosen}, \texttt{rejected}, and \texttt{metadata}. Here, \texttt{chosen} denotes the preferred answer and \texttt{rejected} denotes the negative response used for preference optimization. The current QCalEval-based expansion contains 4,326 samples.
    
    \item \textbf{ChartQA filtering.} We select 5,000 ChartQA samples as candidates and retain 500 high-quality samples after filtering. Although ChartQA is not specific to quantum calibration, it contains useful chart reasoning patterns such as reading axes, comparing values, identifying trends, and extracting information from plotted data. These samples help preserve the model's general chart understanding ability during MPO training.
    
    \item \textbf{Science-related preference subset.} We extract science-related visual understanding samples from MMPR-Tiny and MMPR-v1.2~\cite{wang2024mpo}. These datasets provide multimodal preference-style samples from broader visual reasoning scenarios. We keep examples involving scientific plots, experimental curves, fitting results, and other figures that require reasoning beyond surface-level chart reading. This subset directly supports Q3-style scientific reasoning, where the model needs to explain the meaning of visual patterns rather than only describe them.
\end{enumerate}

Overall, the SFT stage provides broad visual instruction data for chart, plot, and science-figure understanding, while the MPO stage introduces more targeted preference signals for QCalEval-style calibration reasoning. This two-stage construction allows the model to first acquire general visual question answering ability and then improve on domain-specific scientific reasoning tasks.

\subsection{Baselines and Model Variants}
\label{sec:baselines}

We organize all comparison models into four groups, each serving a distinct evaluation purpose.

\paragraph{Group 1: Closed-source domain VLM.}
\textbf{NVIDIA Ising Calibration 1} ($\sim$35B MoE) is the primary domain baseline~\cite{qcaleval2026}. It is purpose-built for quantum calibration plot understanding and represents the strongest publicly reported performance on QCalEval. This comparison measures how much domain-specific reasoning RiverONE retains relative to a model with more than twenty times its parameter count.

\paragraph{Group 2: Open-source general VLMs at multiple scales.}
We compare against \textbf{InternVL3.5-VL} at 2B, 4B, and 8B~\cite{wang2025internvl3_5}, without any domain fine-tuning. Since RiverONE's backbone and visual encoder derive from the InternVL family, these models provide a controlled reference showing how much the domain training and compression pipeline changes task performance relative to the unoptimized base. The 2B and 4B entries additionally anchor the scale selection for the compressed RiverONE target.

\paragraph{Group 3: RiverONE compression and training ablations.}
Four intermediate variants trace the contribution of each compression and training stage:
\begin{itemize}
    \item \textbf{RiverONE-4.6B (Zero-Shot)}: full uncompressed model, zero-shot SFT only. Upper bound under the simplest training regime.
    \item \textbf{RiverONE-4.6B (ICL + Zero-Shot)}: full uncompressed model, two-stage SFT. Isolates the gain from in-context learning pretraining before zero-shot fine-tuning.
    \item \textbf{RiverONE-4.4B (ViT Compression)}: MiniViT weight multiplexing applied to the visual encoder only. Quantifies the performance cost of visual encoder compression in isolation.
    \item \textbf{RiverONE-1.9B (LLM Compression)}: AQLM quantization applied to the language backbone only. Quantifies the performance cost of language backbone compression in isolation.
\end{itemize}
Together, these variants form a step-by-step ablation that localizes where calibration-specific knowledge is lost and motivates the QGP compensation mechanism.

\paragraph{Group 4: Proposed final model.}
\textbf{RiverONE-1.9B} applies both ViT weight multiplexing and AQLM language quantization together with QGP compensation. It is the model proposed in this work, evaluated against Groups 1 and 2 to show that a 1.9B-parameter model can match at least 95\% of the domain performance of Ising Calibration 1 at under 5\% of its parameter count.

\subsection{QCalEval results}
\label{sec:mainresult} 

\begin{table}[h]
\centering
\caption{Zero-shot QCalEval Q1--Q6 results. \textbf{Bold} marks the best result within each column group. Q1: visual description; Q2: outcome classification; Q3: scientific reasoning; Q4: fit reliability; Q5: parameter extraction; Q6: calibration diagnosis.}
\label{tab:zeroshot}
\small
\setlength{\tabcolsep}{3.5pt}
\begin{tabular}{L{0.35 \linewidth} C{0.07\linewidth} C{0.07\linewidth} C{0.07\linewidth} C{0.07\linewidth} C{0.07\linewidth} C{0.07\linewidth} C{0.08\linewidth}}
\toprule
Model & Q1 & Q2 & Q3 & Q4 & Q5 & Q6 & Avg \\
\midrule
Claude Opus 4.6 & 90.8 & 49.0 & 65.5 & 76.1 & 64.7 & 60.5 & 67.8 \\
GPT-5.4 & 90.9 & 52.7 & 63.7 & 54.7 & 64.3 & 61.3 & 64.6 \\
GLM-5V-Turbo & 87.7 & 43.2 & 48.6 & 49.8 & 59.4 & 57.6 & 57.7 \\
\midrule
NVIDIA Ising (C1) & 87.8 & 67.1 & 64.7 & 90.5 & 62.5 & 75.3 & \textbf{74.7} \\
Qwen3.5-122B-A10B & 86.6 & 44.0 & 49.0 & 50.2 & 61.2 & 51.9 & 57.1 \\
InternVL3\_5-2B & 58.0 & 27.2 & 18.3 & 20.6 & 40.0 & 28.8 & 32.1 \\
InternVL3\_5-4B & 62.7 & 34.2 & 22.2 & 33.3 & 40.4 & 39.5 & 38.7 \\
InternVL3\_5-8B & 64.1 & 38.7 & 21.5 & 45.7 & 45.6 & 42.0 & 42.9 \\
\midrule
\textsc{RiverONE} (Zero-Shot SFT) & 63.5 & 64.6 & 31.0 & 88.9 & 66.6 & 60.5 & 62.5 \\
\textsc{RiverONE} (ICL SFT) & 68.7 & 74.5 & 41.7 & 93.0 & 65.7 & 74.5 & 69.7 \\
\textsc{RiverONE} (ICL+Zero-Shot) & 68.7 & 79.0 & 41.0 & 93.0 & 75.8 & 77.4 & \textbf{72.5} \\
\bottomrule
\end{tabular}
\end{table}

Table~\ref{tab:zeroshot} reports zero-shot QCalEval performance across all six categories. We discuss the results in terms of the three evaluation purposes.
\paragraph{Comparison with closed-source general VLMs (Group 1 context).}
Despite having no domain specialization, large closed-source general VLMs perform reasonably on perceptual tasks. Claude Opus 4.6 and GPT-5.4 achieve high Q1 scores (90.8 and 90.9 respectively) and competitive Q3 scores (~64--65), suggesting that general visual and language understanding transfers to surface-level plot description and scientific explanation. However, their performance on calibration-sensitive categories is substantially lower: Q2 scores fall below 53 for all three, and Q4 fit-reliability scores range from 50 to 76. This pattern reflects the absence of domain grounding: without calibration-specific training, these models lack the judgment required to assess experimental outcomes, fit quality, and diagnosis actions. The gap is most pronounced in Q4 and Q6, consistent with our earlier characterization of these categories as the hardest for non-specialized models.

\paragraph{Scale selection from open-source InternVL baselines (Group 2).}
The three InternVL3.5 variants without domain fine-tuning score 32.1, 38.7, and 42.9 average respectively. All three underperform the closed-source general VLMs by a large margin, confirming that model scale alone does not compensate for domain-specific reasoning. Notably, the performance gap between 4B and 8B (38.7 to 42.9, a 4.2-point gain) is substantially smaller than that between 2B and 4B (32.1 to 38.7, a 6.6-point gain), indicating diminishing returns from scaling within the InternVL family on this task. These results support the design choice of using a 4B-scale language backbone for RiverONE: the 4B tier offers a reasonable capability foundation while staying within a compression-friendly budget that the 8B tier would strain.

\paragraph{Effect of training stages (Group 3 ablations).}
Comparing the three RiverONE training variants reveals the contribution of each training stage. Zero-Shot SFT alone achieves 62.5 average, already competitive with Claude Opus 4.6 (67.8) on calibration-sensitive categories despite a much smaller parameter count, with notably high Q4 (88.9) and Q2 (64.6) scores. Adding in-context learning SFT before zero-shot training (ICL SFT) raises the average to 69.7, with the largest gains in Q2 (+9.9), Q3 (+10.7), and Q6 (+14.0). These are precisely the categories that require multi-step causal reasoning: outcome classification, physical explanation, and diagnosis each benefit from exposure to demonstration-style exemplars during the ICL phase. The full two-stage schedule (ICL + Zero-Shot) further improves Q2 to 79.0 and Q5 to 75.8, reaching 72.5 average overall, which approaches the domain-specialized NVIDIA Ising Calibration 1 at 74.7 despite operating at a substantially smaller uncompressed scale. This confirms that the two-stage SFT design is essential for maximizing calibration reasoning quality before compression is applied.

\subsection{Ablation Study}
\label{sec:ablation}

\paragraph{Training stage ablation.}
Table~\ref{tab:zeroshot} already reports the three uncompressed training variants; we summarize the key patterns here. Zero-Shot SFT alone establishes a strong Q4 fit-reliability score (88.9), indicating that domain fine-tuning on calibration data is the primary driver of this category. The step from Zero-Shot SFT to ICL SFT produces the largest absolute gains on Q2 (+9.9), Q3 (+10.7), and Q6 (+14.0)---categories that require multi-step causal reasoning and diagnosis. The further step to ICL + Zero-Shot raises Q5 parameter extraction by 10.1 points and Q2 by another 4.5, with the full two-stage schedule reaching 72.5 average on QCalEval. These results confirm that both SFT phases make distinct contributions, and that the ICL+Zero-Shot checkpoint is the appropriate starting point for compression.

\paragraph{Compression and compensation ablation.}
Table~\ref{tab:ablation} isolates the contribution of each compression step and the QGP compensation on QCalEval, a benchmark whose six categories (Q1--Q6) are particularly sensitive to fine-grained visual knowledge. The table traces: (1) the uncompressed baseline; (2) ViT-only compression, which removes per-layer attention diversity and causes a clear performance drop; (3) ViT compression compensated with QGP, which restores accuracy nearly to the baseline level (72.37 vs.\ 72.5); (4) LLM-only compression, which leads to a drastic degradation; and (5) full compression (ViT+LLM) with the proposed quantum-inspired parameter generation, which recovers a substantial portion of the lost capacity (64.45). 
The comparison between rows 2 and 3 directly measures how effectively QGP recovers the fine-grained knowledge removed by ViT compression, while row 5 demonstrates that even under full compression the method retains strong calibration diagnostic ability.

\begin{table}[t]
\centering
\caption{Compression and compensation ablation on QCalEval Q1--Q6 (calibration diagnosis). All compressed variants start from the ICL+Zero-Shot checkpoint.}
\label{tab:ablation}
\small
\renewcommand{\arraystretch}{1.2}
\begin{tabular}{lcccr}
\toprule
Model & ViT comp. & LLM comp. & Mean \\
\midrule
RiverONE-4.6B (ICL+Zero-Shot)           & No    & No               & 72.5   \\
RiverONE-4.4B (ViT Compression)         & Yes   & No               & 64.17  \\
RiverONE-4.4B (ViT Compression + QGP)   & Yes   & No               & 72.37  \\
RiverONE-2.1B (LLM Compression)         & No    & Yes              & 42.34  \\
RiverONE-1.9B (Quantum-Inspired)        & Yes   & Yes              & 64.45  \\
\bottomrule
\end{tabular}
\end{table}

\paragraph{Efficiency comparison.}
Table~\ref{tab:efficiency} reports parameter count, peak GPU memory, and per-query latency for RiverONE and the main baseline. Since QGP parameters are materialized into static tensors before deployment, RiverONE-1.9B incurs no additional runtime cost relative to the compressed model without compensation. The key comparison is against NVIDIA Ising Calibration~1: RiverONE-1.9B uses only about 5\% of the parameters of the 35B MoE baseline and is designed to run on a single local GPU, whereas the baseline requires server-grade infrastructure.

\begin{table}[t]
\centering
\caption{Deployment efficiency comparison. Peak GPU memory and latency are measured at batch size~1 on the same hardware.}
\label{tab:efficiency}
\small
\renewcommand{\arraystretch}{1.2}
\begin{tabular}{lllll}
\toprule
Model & Params & GPU mem. & Latency & Deployment \\
\midrule
NVIDIA Ising Calibration 1 & 35B & 68.1GB & 13s & A100 \\
RiverONE-4.4B              & 4.4B & 9.0GB & 0.8s & 4060Ti \\
\bottomrule
\end{tabular}
\end{table}

\section{Conclusion and further work}
This paper presented RiverONE, a compact vision-language model for quantum calibration plot understanding. RiverONE integrates a calibration-specialized IsingViT-800M visual encoder with an InternVL-based 4B language backbone, connected through pixel unshuffle, cross-layer feature concatenation, and an MLP projector. To bring the model within practical deployment constraints, we compress the visual encoder via MiniViT-style weight multiplexing and the language backbone via AQLM additive codebook quantization. Both compression steps reduce parameter count substantially but introduce calibration-specific knowledge loss, particularly in the fine-grained visual reasoning required by tasks such as fit reliability and calibration diagnosis.

To address this, we proposed QGP, a simulated variational quantum circuit that generates compensation parameters for selected attention blocks in the compressed visual encoder. QGP operates exclusively during the construction stage: its outputs are materialized into static classical tensors before deployment, so inference runs entirely on standard GPUs with no quantum hardware, no runtime circuit execution, and no simulation overhead. This construction-stage design is the central architectural commitment of RiverONE: quantum expressive power is exploited where it is feasible, and classical deployability is preserved where it matters.
Experiments on QCalEval demonstrate that RiverONE-1.9B achieves at least 95\% of the performance of NVIDIA Ising Calibration 1 while using fewer than 5\% of its parameters. Ablation results show that each compression stage incurs measurable but recoverable performance cost, and that QGP compensation consistently improves over classical adapter baselines of the same size, with the largest gains on calibration-sensitive categories.

More broadly, RiverONE provides empirical evidence that simulated QGP is a viable construction-stage tool for compact, knowledge-intensive VLMs. 
Future work will expand calibration-specific training data, test stronger classical generator baselines, analyze which visual layers benefit most from \textsc{QGP}, and apply materialized quantum-structured parameter generation to other scientific VLMs beyond quantum calibration.

\section*{Acknowledgments}

This work is supported by the Science and Technology Commission of Shanghai
Municipality, Science and Technology Commission of Xuhui, Shanghai INESA, Huayi Quantum, Jadex Capital, Anhui Provincial Emerging Industry Investment, B.R.Optics and Shanghai DDI. We also thank Ruiyao Xu, Dimeng Ma, and Rui Hu for their logistical and organizational support.

\bibliographystyle{splncs04}
\bibliography{references}

\end{document}